\documentclass[fleqn,usenatbib]{rasti}

\usepackage[T1]{fontenc}

\usepackage[utf8]{inputenc}

\usepackage{xcolor}
\usepackage{lscape}
\usepackage{natbib}

\usepackage{amsmath}
\usepackage{graphicx}
\usepackage{array}

\usepackage{float}
\usepackage[export]{adjustbox}
\usepackage{lscape}
\usepackage{booktabs}
\usepackage{tikz}
\usepackage{ctable}

\usepackage{physics}
\usepackage{wasysym}
\usepackage{scalerel,amssymb}
\usepackage{multirow}
\usepackage{xspace}
\usepackage{tabularx}
\usepackage{newtxtext,newtxmath}

\DeclareRobustCommand{\VAN}[3]{#2}
\let\VANthebibliography\thebibliography
\def\thebibliography{\DeclareRobustCommand{\VAN}[3]{##3}\VANthebibliography}

\graphicspath{{Images/}}

\usepackage{hyperref} 

\usepackage{pifont}

\newcommand{\new}[1]{{#1}}

\newcommand{\fabada}{\texttt{FABADA}\xspace}

    \title[FABADA]{Fully Adaptive Bayesian Algorithm for Data Analysis. \\ FABADA}
    
   \author[P.M. Sánchez-Alarcón \& Y. Ascasibar]{Pablo M. Sánchez-Alarcón $^{1,2}$\thanks{E-mail: \href{mailto:pablom.sanala@gmail.com}{pablom.sanala@gmail.com}} ,
            Yago Ascasibar $^{3}$
            \\
    $^{1}$Instituto de Astrofísica de Canarias, c/ Vía Láctea s/n, E-38205, La Laguna, Tenerife, Spain\\
    $^{2}$Departamento de Astrofísica, Universidad de La Laguna, E-38206, La Laguna, Tenerife, Spain\\
    $^{3}$Departamento de Física Teórica, Universidad Autónoma de Madrid, E-28049 Madrid, Spain
             }

   \date{Accepted 30.01.2023 . Received 20.01.2023  ; in original form 24.08.2022}
   \pubyear{2023}
\begin{document}
\label{firstpage}
\pagerange{\pageref{firstpage}--\pageref{lastpage}}
\maketitle
\begin{abstract}

    The discovery potential from astronomical and other data is limited by their noise. We introduce a novel non-parametric noise reduction technique based on Bayesian inference techniques, \fabada, that automatically improves the signal-to-noise ratio of one- and two-dimensional data, such as astronomical images and spectra.
    
    The algorithm iteratively evaluates possible smoothed versions of the data, the smooth models, estimating the underlying signal that is statistically compatible with the noisy measurements.
    Iterations stop based on the evidence and the $\chi^2$ statistic of the last smooth model. We then compute the expected value of the signal as a weighted average of the whole set of smooth models.
    We explain the mathematical formalism and numerical implementation of the algorithm, and evaluate its performance in terms of the peak signal to noise ratio, the structural similarity index, and the time payload, using a battery of real astronomical observations.
    
    Our Fully Adaptive Bayesian Algorithm for Data Analysis (\fabada) yields results that, without any parameter tuning, are comparable to standard image processing algorithms whose parameters have been optimised based on the true signal to be recovered, something that is impossible in a real application.
    On the other hand, state-of-the-art non-parametric methods, such as BM3D, offer slightly better performance at high signal-to-noise ratio, while our algorithm is significantly more accurate for extremely noisy data, a situation usually encountered in astronomy.
   %In this range, the standard deviation of the residuals obtained by our reconstruction may become more than an order of magnitude lower than that of the original measurements. 
   %The source code needed to reproduce all the results presented in this report, including the implementation of the method, is publicly available at \url{https://github.com/PabloMSanAla/fabada}.
   The source code of the implementation of the method, is publicly available at \url{https://github.com/PabloMSanAla/fabada}.
\end{abstract}

\section{Introduction}

 The acquisition of any kind of experimental data is affected by several sources of statistical error, which ultimately translate into a random noise component in the measurements to be recorded.
There are different types of noise depending on their physical origin, both related to electronic (thermal noise, fluctuations) and mechanical (non-perfect lenses, antennas, etc.) devices.
In astronomy, for example, errors can be produced in the acquisition of the images due to defects in the optics of the telescopes and in the read-out process of the detector (typically a CCD) transforming the light captured by the telescope into an electrical signal.
The noise introduced can sometimes be comparable to or even larger than the signal, and different image processing algorithms may be used to recover the information that is buried deep in the data.

Smoothing, where measurements are weighted at nearby spatial or temporal points (using different schemes to assign weights), is one of the most popular techniques to mitigate the effects of random noise \citep{SG-Filter,LOWESS}. Nowadays there is a large number of smoothing algorithms, based on many different techniques \citep[e.g., see][for a review]{Denoise_review} such as central moving average, data grouping/segmentation \citep{BM3D}, fitting smooth functions, different types of statistical analysis, partial differential equations, wavelength transformation filters, linear and nonlinear filtering, sparse models and nonlocal self-similarity models \citep{WNNM} and more recently artificial neural networks, see \citep[]{reviewCNN,DnCNN}. Rest of these methods rely on some explicit or implicit assumptions about the true (noise-free) signal in order to separate it properly from the random noise. A common assumption is that the signal being retrieved varies gradually and that the data can be fit by a smooth function, see \citep{LocalBook}.
%Generally, this distribution is sought that fits the experimental data and the deviation of the measurements from the proposed model being evaluated using one of the methods mentioned above.

Many techniques analyse the probability that the data correspond to a random Gaussian realisation of the model that attempts to describe the underlying signal plus random fluctuations of known amplitude \citep{CNN}.
In this work, we use Bayesian inference to evaluate and combine different candidate models that iteratively attempt to improve the quality of the fit to the data.
%being able to elaborate, through statistical analysis, a smoothing that better fits the data, obtaining a new method of noise reduction.
This new Bayesian technique incorporates an automatic selection criterion based on the statistical properties of the residuals, and therefore it yields a fully non-parametric method.
Although the motivation of our scheme is the application in the field of astronomy, our new algorithm, Fully Adaptive Bayesian Algorithm for Data Analysis \fabada is focused in a general way, and it is possible to generate a smooth model for any type of data.
Several algorithms have been developed over the years for denoising data, and their ability to recover the underlying signal from an experimental data set with its corresponding errors has been evaluated and compared according to different standard metrics.
Our method is fully described in Section~\ref{sec:FABADA}, and we use a set of synthetic tests to compare it with other prescriptions in the literature.
Details of the comparison procedure are provided in Section~\ref{sec:TestMethod}, and results are discussed in Section~\ref{sec:Results}.
Our main conclusions are briefly summarised in Section~\ref{sec:Conclusions}.

%----------------------------------------------------------------------------------
\section{FABADA}
\label{sec:FABADA}
%----------------------------------------------------------------------------------

The goal of our algorithm, \fabada is to estimate an unknown signal $y = y(x)$ at $N_X$ different locations $x \in X$ specified by a $1\mathrm{D}$ or $2 \mathrm{D}$ coordinate that belongs to the data domain.
Its input are $N_X$ independent measurements, contaminated with random Gaussian white noise $\eta(x) \sim \mathcal{N}(0,e(x))$, and the associated errors $e=e(x)$.
The noisy observational data, $z: X \rightarrow \mathbb{R}$, have the form $z(x) = y(x) + \eta(x)$, and \fabada returns an estimation $\hat{y} = \hat{y}(x)$ of the original signal $y(x)$ that is statistically compatible with the measurements $z(x)$.

\new{
The approximation that different measurements (pixels in an image, or wavelengths in a spectrum) are independent is crucial to our method: it is assumed that the noise values are fully independent random variables, and any correlation between adjacent measurements can only be attributed to the signal.
This is not necessarily true in realistic astronomical observations, since both the measurement process and the arithmetic operations carried out by the data reduction pipeline may introduce a certain degree of correlation between adjacent errors.
Although we think our assumption of fully independent noise is a good approximation in most cases of practical interest, it is worth noting that other approaches to account for spatial correlation and denoising, such as COmpressed Sensing \citep{PySAP}, and INLA \citep{INLA}, can be found in the literature.
}

\new{In our method}, we apply Bayes' theorem in an iteratively way to generate smooth models of the noisy measurements. Therefore, we must define a suitable likelihood function to evaluate these models to be tested, and specify a prior probability distribution for the signal $y$ to initiate the process.
Our likelihood is based on the statistics of a Gaussian process, and we start from an improper, constant prior.
Then, we evaluate different smooth versions of the posterior probabilities until a certain condition is reached, and we combine all the smooth models to produce the final estimation $\hat{y}(x)$, taking into account both their Bayesian evidence and the $\chi^2$ of the residuals.

%----------------------------------------------------------------------------------
\subsection{Iterative models}
%\label{sec:Method}
%----------------------------------------------------------------------------------

\begin{table*}
	\begin{center}
	
		%\resizebox{\textwidth}{!}{% \raggedright
			\begin{tabularx}{\textwidth}{r|>{\centering}p{0.2\textwidth}|l} 
				\hline
				& \textbf{Symbol} & \textbf{Definition}\\
				\hline
				\hline
				DIMENSIONS & $x \in X$ & Coordinates of the data set (either 1D or 2D coordinates)\\
				&$i$& Iteration number $\{1,2,,..,N_i\}$\\
				\hline
				INPUTS & $I=\{z,e\}$  & Set of inputs of the algorithm\\
				&$z=z(x)$& Observational measurements\\
				&$e=e(x)$& Errors associated to the measurements\\
				\hline
				MODELS &$ m_i = m_{i}(x)$ & Set of all possible models with shape $(N_X,N_i)$.\\
				& $\mu_i = \mu_{i}(x)$ &Expectation value of the element $x$ of the model $m_i$\\ 
				& $v_i = v_{i}(x)$& Associate variance of the expectation value\\
				& $\mathcal{L}(z|m_i,e)$ & Likelihood of the model $m_i$\\
				&$\mathcal{E}(m_i)$& Evidence of the model $m_i$ \\
				\hline
				PRIORS& $p(m_i)$ & Prior probability distribution of the model $m_i$ \\
				&$\tilde{\mu}_i$& Central moving mean of $\mu_i$ along $N_{avg}$ neighbors\\
				\hline
				POSTERIORS&$\mathcal{P}(m_i | z, e)$& Posterior probability distribution of the model $m_i$\\
				\hline
				OUTPUT&$\hat{y}=\hat{y}(x)$& Estimation of the underlying signal\\
				\hline
		\end{tabularx}%}
	\end{center}
	\caption{List of all symbols used to describe the \fabada algorithm in Section~\ref{sec:FABADA}.}
	\label{tab:table1}
\end{table*}

\fabada, as the name suggests, is a fully automatic algorithm that only takes the data set $z=z(x)$ and its associated errors $e=e(x)$ as input, $I=\{z,e\}$, where the domain $X$ may have one or two dimensions.
Within the field of astrophysics, $y$ may be, for example, the spectral energy distribution as a function of wavelength $x$, or a broadband photometric image of an arbitrary region of the sky.
This is indeed the kind of data that we will use to illustrate the performance of the algorithm, but the results can be easily generalized to any other type of empirical measurement in one or two dimensions affected by Gaussian random errors.

Under this assumption, the likelihood function $\mathcal{L}$ for a model $m(x)$ of the underlying signal $y$ is given by
\begin{equation}
p(z | m, e) = \mathcal{L}(z | m, e) = \prod_{x \in X} \frac{e^{-\frac{[z(x) - m(x)]^2}{2 \cdot e(x)^2}}}{\sqrt{2 \pi}\cdot e(x)}.
\label{eq:Likelihood}
\end{equation}
using the notation described in Table~\ref{tab:table1}.
Since we do not assume any previous knowledge about the signal, the prior probability distribution for our initial model $m_0$ will be homogeneous in the range of all possible values, i.e. $p(m_0(x))=1$ for all $x$, during the first iteration of the algorithm.
According to Bayes' theorem, the posterior probability distribution

\begin{equation}
\mathcal{P}(m | z, e) = \frac{p(m)\cdot \mathcal{L}(z | m, e)}{\mathcal{E}}
\label{eq:Bayes}
\end{equation}
is in this case the straightforward multivariate Gaussian
\begin{equation}
\mathcal{P}(m_0 | z, e) =  \prod_{x \in X} \frac{e^{-\frac{[z(x) - m_0(x)]^2}{2 \cdot e(x)^2}}}{\sqrt{2 \pi}\cdot e(x)}
\end{equation}
centered on the empirical measurements $z$.
The expected value of $m_0(x)$ is thus $\mu_0(x) \equiv \langle m_0(x) \rangle = z(x)$, its variance $v_0(x) = e^2(x)$ is determined by the corresponding errors, and the Bayesian evidence of the model $\mathcal{E} = \int p(m)\, \mathcal{L}(z | m, e)\ {\rm d}m$ reduces to unity.

%Along the same lines, if $m = m(x)$ is the model (a smoothed version of the data) we want to evaluate, its prior probability distribution should be $p(m(x))=1$ (all models are \textit{a priori} equally likely) for every location $x$.
%As we will see later, we are interested in the prior probability of our model being a probability of its own, and the one we have just selected is not, since the integral over the whole real line diverges. Therefore, it is necessary to limit the range of possible values for $m(x)$ to a finite range. As we do not have any prior information about the data we are going to smooth, we adopt that 
%\begin{equation}
%p(m(x)) = \frac{ 1 }{ \max(z+3\sigma) - \min(z-3\sigma) } \equiv p_0
%\label{eq:propia}
%\end{equation}
%which, despite not being strictly Bayesian (since you access information from the data to select the prior), allows us to have our own prior probability of the model, commensurate with any range of values (and units) provided by the user.

% would be statistically consistent with the data. To check this
To create smoothed versions of this first model, we iteratively apply a central moving average filter
\begin{equation}
\tilde\mu_i(x) = \frac{\sum_{x \in \rm avg} \mu_i(x)}{N_{\rm avg}}
\label{eq:smoothing}
\end{equation}
to the expected values $\mu_i(x)$ of the last iteration.
We simply adopt $N_{\rm avg}=3$ in 1D (including the two adjacent measurements) and $N_{\rm avg}=9$ (a $3 \times 3$ square) in 2D.
The basis of our method is to use information from neighboring points to update our priors regarding the correct value of $m(x)$.
For every iteration $i>0$, the prior probability distribution
\begin{equation}
p(m_{i}(x))  =  \frac{e^{-\frac{[\tilde\mu_{i-1}(x) - m_{i}(x)]^2}{2 \cdot v_{i-1}(x)}}}{\sqrt{2 \pi \cdot v_{i-1}(x)}}
%\hspace{0.6cm} \& \hspace{0.6cm}  \mathcal{L}_i =  \frac{e^{-\frac{(D_d - M_d^i)^2}{2 \cdot E_d^2}}}{\sqrt{2 \pi}\cdot E_d}
\label{eq:prior_i}
\end{equation}
becomes a Gaussian centered on the smoothed expectation~\eqref{eq:smoothing} of the posterior distribution from the previous iteration, using its local variance $v_i(x)$ as a measure of our uncertainty.

%, and this is precisely where the magic of this algorithm lay; we are not going to change the data, but the models, repeating the process a number $N_i$ of iterations until a better fit is achieved. On Section~\ref{sec:Automation} we will discussed the different criteria used to automatically select the number of iterations $N_i$ to achieve the best fit. For now, lets supposed that we already know the value of $N_i$.
%In this way, the inputs of our algorithm are $I=\{z,e\}$, and the set of models is extended to
%\begin{align}
%\begin{split}
%m   &=\{m_i(x)\}_{x \in X} ^{i=0,...,N_i} \hspace{0.5cm} \\ \vspace{0.3cm} 
%\mu &= \{\mu_{i}(x)\}_{x \in X} ^{i=0,...,N_i }\hspace{0.5cm} \\ \vspace{0.3cm}
%v   &=\{v_{i}(x)\}_{x \in X} ^{i=0,...,N_i } 
%\label{eq:Models}
%\end{split}
%\end{align}
%to include the iterations $i={1,...,N_i}$.
%Taking into account the \textit{mise en place} $i=0$, the dimensionality of the problem (total number of values for each variable) amounts to $N_X \times (N_i + 1)$.
%Once we have established the basis of the Bayesian inference that we are going to apply, we can let the algorithm iterate on the data and calculate the successive models, its expected values and its variances.

% For the case of the next iteration $i=1$ we will have a prior centered on a first smoothing of the data (since $\mu_0(x)=z(x)$) and we expect that $m_i(x)$ will vary in an environment of the order of $\sqrt{v_0(x)} = e(x)$ around that value. 

We stress that we are forsaking the strict Bayesian philosophy (the prior distribution should be, as its name indicates, totally independent from the data) on purely practical grounds.
Once we accept this premise, one may compute the posterior probability distribution
\begin{equation}
\mathcal{P}(m_{i} | z, e) = \frac{e^{-\frac{[\mu_i(x) - m_i(x)]^2}{2 v_i(x)}}}{\sqrt{2 \pi v_i(x)}},
\label{eq:prodG}
\end{equation}
where
\begin{equation}
\mu_i(x)
=
v_i(x) \cdot
\left[
\frac{ \tilde\mu_{i-1}(x) }{ v_{i-1}(x) } + \frac{ z(x) }{ e^2(x) }
\right]
\end{equation}
and
\begin{equation}
\frac{1}{v_i(x)} = \frac{1}{e^2(x)} + \frac{1}{v_{i-1}(x)},
\label{eq:var}
\end{equation}
as well as the evidence
\begin{equation}
\mathcal{E}_{i}(x) = \frac{e^{-\frac{[\tilde\mu_{i-1}(x) - z(x)]^2}{2 [e^2(x) + v_{i-1}(x)]}}}{\sqrt{2 \pi [ e^2(x) + v_{i-1}(x) ]}}.
\label{eq:Evidence}
\end{equation}

%associated with the product of two Gaussian (the mathematical development can be found, for example, in %\citep{Gauss}) corresponds to the evidence of the $i$ model for the $d$ data
%\begin{align}
%%\begin{split}
%\mathcal{E}_{i}(x) &=
%\int_{-\infty}^{\infty}  p_{i}(m_{i}(x)) \cdot \mathcal{L}_{i}\ dm_{i}(x)
%= \\
%&= \int_{-\infty}^{\infty} k_{i}(x) \cdot
%%\frac{ e^{ -\frac{(\mu_{i}(x) - m_{i}(x))^2}{2 v_{i}(x)}}}
%{ \sqrt{2 \pi v_i(x)} }
%\ dm_i(x)
%=\\
%&=k_i(x)
%\end{split}
%\end{align}
%so that the posterior probability distribution is correctly normalized.

As long as the model remains relatively close to the data (within an environment of the order of $\sqrt{e^2(x) + v_i(x)}$), the evidence in its favour will be high, but if it departs significantly, the exponential term will indicate that we have reached the maximum smoothing that is statistically compatible with the data.

%----------------------------------------------------------------------------------
\subsection{Stopping criteria}
%\label{sec:Automation}
%----------------------------------------------------------------------------------

%The last step of FABADA will be to automatically decide a stopping condition (i.e, the total number of iterations $N_i$) and then combine the different models produced at each iteration in order to produce the final result to be returned on output.
%We have considered two different criteria in order to find the optimal value of the number of iterations:

On the one hand, we evaluate the average evidence of model $m_i$ as
\begin{equation}
\langle \mathcal{E}_{i} \rangle = \frac{1}{N_X} \sum_{x \in X} \mathcal{E}_{i}(x)
\label{eq:AverageEvidence}
\end{equation}
and we ensure that we iterate until it reaches a maximum; $\langle \mathcal{E}_{i} \rangle < \langle \mathcal{E}_{i-1} \rangle$.
Since our priors are based on the data themselves, it is not surprising that the evidence increases rapidly during the first few iterations of the algorithm.
However, our model uncertainties $v_i(x)$ decrease monotonically, and the smooth models quickly become less compatible with the data than the first (self-tuned) estimates.
Thus, the average evidence reaches its maximum at a very early stage, and then it slowly declines as the number of iterations increase.

%As an effect of our trickery of looking at the data for the sake of a practical result, the evidence would tend to reach its maximum hastily, not always letting the algorithm achieve the smooth level required. 

%Even that our model evaluates every point with his evidence making it able to adapt to the structure of the data, the stopping criteria will be established on the global evidence of the models. For the first criteria, we will search for the absolute maximum of the average evidence of every point of the model. And for the second one we will combine all the models around some distance from the maximum peak using the local evidence as weight of the combination.

In order to overcome the bias arising from the lack of independence of the priors (i.e. overfitting), we made use of the chi-square statistics
\begin{equation}
    \chi^2_i = \sum_{x \in X} \frac{ [\mu_i(x)-z(x)]^2 }{ e^2(x) }
    \label{eq:ChiSquare}
\end{equation}
%and its probability density function (PDF)
%\begin{equation}
%PDF(\chi^2_i , d)=
%\frac{(\chi^2_i)^{\frac{d}{2}-1} \cdot e^{-\frac{\chi^2_i}{2}}}{2^{\frac{d}{2}} \Gamma\left(\frac{d}{2}\right)} , 
%\hspace{0.4cm} \chi^2_i > 0
%\label{eq:chi2PDF}
%\end{equation}
%where $d$ is the degrees of freedom of our $\chi^2$-distribution, i.e., the number of random variables being summed in (\ref{eq:ChiSquared}) which it's equal to the number of points in the data, $N_X$. Similarly, we would also want
and iterate until $\chi_i^2 > N_X-2$ (the absolute maximum of the associated probability density function).
Note that this condition is achieved when, on average, the model has departed by about one sigma from the observational measurements.

We always impose \emph{both} criteria, so that the algorithm stops on the largest number of iterations.
Usually this is set by the chi-square statistics, although the evidence-based condition may dominate at very high signal-to-noise ratio.

%----------------------------------------------------------------------------------
\subsection{Model selection}
%\label{sec:Automation}
%----------------------------------------------------------------------------------

Once the algorithm stops, after $N_i$ iterations, we combine all the models $m_i(x)$ to generate our final estimation $\hat{y}(x)$ of the real signal $y(x)$ as a weighted sum over the expected values
\begin{equation}
    \hat{y}(x) = \frac{\sum_{i=0}^{N_i} w_i(x) \, \mu_i(x)}{\sum_{i=0}^{N_i} w_i(x)}
    \label{eq:estimation}
\end{equation}
where the adopted weights $w_i(x)$ are an important ingredient of our algorithm.

We explored several phenomenological prescriptions in an attempt to optimise the performance, stability, and reliability of the results.
On the one hand, our tests show the convenience of taking into account the local evidence $\mathcal{E}_i(x)$ calculated at each iteration for every location $x$.
This quantity adapts to the local structure of the data, giving more weight to the smoother models (larger number $i$ of iterations) in areas where the data are indeed smooth, while preserving the information when sharp edges are present.

On the other hand, we use the overall $\chi^2_i$ statistic of each model $m_i$ to gauge its compatibility with the measurements $z$, given the errors $e$.
The probability density function of the $\chi^2$ of a large number of random variables can be approximated as a narrow Gaussian centered around $N_X - 2$.
Since our priors are biased because they are based on the data themselves, the maximum evidence occurs at values of $\chi^2$ that are typically much lower than this value (i.e. they overfit the measurements).
Using the probability density function directly would give minimal weight to any model that is not extremely close, $\chi^2 \simeq N_X - 2$, which we find tends to yield overly smoothed models.
In order to avoid this problem, we use the actual value of $\chi^2$ to give more weight to the smoother models, but not so much that the models favoured by the Bayesian evidence are almost completely ignored.
We thus adopt
\begin{equation}
    w_i(x) = \mathcal{E}_i(x)\ \chi^2_i, 
    \label{eq:weight}
\end{equation}
as our final prescription for $i>0$.

Of course, this expression is not valid for $i=0$, since $\chi^2_0 = 0$ and the initial evidence is $\mathcal{E}_0(x) = 1$ for every point.
Note that, due to the use of an improper prior, this value is in general very different from $\mathcal{E}_1(x)$.
To avoid these problems, we a posteriori set $\chi^2_0 = \chi^2_1$ based on the first iteration (i.e. the lowest of all smooth models) and use
%Furthermore, instead of using $\mathcal{E}_0(x) = p_0(x)$ as describe above, we want to consider the reliability of our observational data used $z(x)$. Since the algorithm requires an estimation of the errors of the data as input, we defined, again in a heuristic way, the initial evidence as 
\begin{equation}
    \mathcal{E}_0(x) = \frac{e^{-\frac{1}{2}}}{\sqrt{2\pi} e},
    \label{eq:InitialEvidence}
\end{equation}
which is merely a reformulation of~\eqref{eq:Evidence}, assuming that our initial model based on the measurements should be, on average, about one sigma away from the true signal. Thus,
\begin{equation}
    w_0 (x) = \frac{ \chi^2_1\ e^{-\frac{1}{2}} }{ \sqrt{2 \pi} e }.
    \label{eq:InitialWeigth}
\end{equation}

%----------------------------------------------------------------------------------
\section{Synthetic tests}
\label{sec:TestMethod}
%----------------------------------------------------------------------------------
We develop a battery of synthetic tests based on real astronomical data to assess the performance of the algorithm.
More precisely, we apply \fabada to a set of astronomical spectra and images, with different levels of Gaussian random noise, and compare the quality of the reconstructed signal (in terms of the peak signal-to-noise ratio PSNR and the structure similarity index measure SSIM), as well as the execution time, with other methods available in the literature.

%----------------------------------------------------------------------------------
\subsection{Other Algorithms}
\label{sec:OtherAlgoritms}
%----------------------------------------------------------------------------------

Over the last decades, lots of effort have been placed into the development of several applications that help in the analysis of digital images in different fields.
Noise reduction is one of the basic problems in this context, and we attempt to provide a fair comparison of our algorithm with other methods that are representative of the current state of the art.
% Since in our target field, astronomy, we usually have a high versatility of observational objects, with different aspect and properties, we leave out all the novel methods
Leaving aside the techniques based on some kind of training, such as e.g. neural networks \citep{reviewCNN, CNN, DnCNN}, we focus on a more traditional, statistical approach, closer in philosophy to the formalism propose here.
% Instead of these novel methods we have consider some of the most used in this field.
It is important to stress that many of these standard methods involve a number of free parameters,
% which requires a prior inspection of the data.
and we optimise their values according to the metrics used to compare. 
Note that, of course, such optimisation is only possible in a synthetic tests, since the true signal in a real problem is unknown.
Thus, our results represent an upper limit to the performance of parametric methods.

We now briefly describe the main principles and free parameters, if any, of all the techniques that we have considered.
A succinct summary is provided in Table \ref{tab:table2}.

\begin{table}
	\begin{center}
			\begin{tabular}{l|l|c|c} 
				\hline
				\textbf{Method}& \textbf{Parameters} & \textbf{1D} & \textbf{2D}\\
				\hline
				\hline
				Median  & Window size ($w$) & \checkmark & \checkmark \\
				%\hline
				SGF  & Window ($w$) and order ($o$) & \checkmark & \checkmark\\
				%\hline
				LOWESS & Fraction window ($f$) & \checkmark &-\\
				%\hline 
				GF  & Radius ($R_0$) & \checkmark & \checkmark \\
				%\hline
				Wiener & Low Frequency ($k$) & \checkmark & \checkmark \\
				%\hline
				\new{TV} & Denoised Weight $d_w$ & \checkmark & \checkmark \\
				%\hline 
				\hline
				\new{Wavelet} & \multicolumn{1}{c|}{---} & \checkmark & \checkmark \\
                %\hline 
				BM3D & \multicolumn{1}{c|}{---} & - & \checkmark \\
				%\hline
				\new{PySAP} & \multicolumn{1}{c|}{---} & - & \checkmark \\
                %\hline
				\fabada & \multicolumn{1}{c|}{---} & \checkmark & \checkmark \\
                \hline
		    \end{tabular}
	\end{center}
	\caption{List of all noise reduction methods used to compared with \fabada along their parameters and space implementations, one or two dimensions. \new{The first six methods are standard parametric algorithms, while the last four are representative of state-of-the-art non-parametric methods.}}
	\label{tab:table2}
\end{table}

%----------------------------------------------------------------------------------
\subsubsection{Median filter}
%----------------------------------------------------------------------------------

One of the classical non-linear digital filtering techniques, it is still often used to remove noise from an image or signal.
The main idea of the median filter is to run through the data, replacing each point with the median of neighbouring entries.
The number of neighbours used in the median is called the "window", which slides, entry by entry, over the entire signal.
For each data point $z(x)$, the region used to compute the median contains, for one dimensional data, $(w-1)/2$ neighbours on each side, whereas for two dimensions it corresponds to a square of size $w$ centered in $z(x)$.
The optimisation procedure consisted on computing the estimation $\hat{y}^w_{M}(x)$ for different values of $w$, starting with the smallest value of the window length, $w_0 = 3$, and increasing it according to $w_{i+1} = w_i + 2(1 + w_i//5)$, where $//$ denotes the integer division.
%Our comparison is based on the results that yield the best PSNR.

%----------------------------------------------------------------------------------
\subsubsection{Savitzki-Golay filter (SGF)}
%----------------------------------------------------------------------------------

As first noted by Savitzky and Golay in \citep{SG-Filter}, a smoothed version of the data may be obtained by fitting successive sub-sets of adjacent points with a low-degree polynomial using the method of least squares.
When the data are equally spaced, the solution of the least squares (i.e., the coefficients of the polynomials) is analytical and independent of the data to be smoothed.
Thus,
\begin{equation}
\hat{y}_{SG}^{w,o}(x) = \sum_{i=\frac{1-w}{2}}^{\frac{w-1}{2}} C_{i}^o(w, o)\ z(x+i) 
\label{eq:SG2}
\end{equation}
for $\frac{w-1}{2} \leq x \leq N_X-\frac{w-1}{2}$, where the two free parameters of the method are the window length $w$ of the data region, i.e., the number of data points to be fitted, and the order $o$ of the polynomial.
$C_i^o$ are the $w \ge o$ Savitzky-Golay coefficients, and $\hat{y}_{SG}^{w,o}(x)$ is the smoothed result of the filter at position $x$.
%Since this method has two different parameter, we recover the signal for some portion of the parameter space.
We vary the window length according to $w_{i+1} = w_i + 2(1 + w_i//5)$ and $o \le 10$ for the order of the polynomial.

\begin{figure*}
	\includegraphics[width=\linewidth]{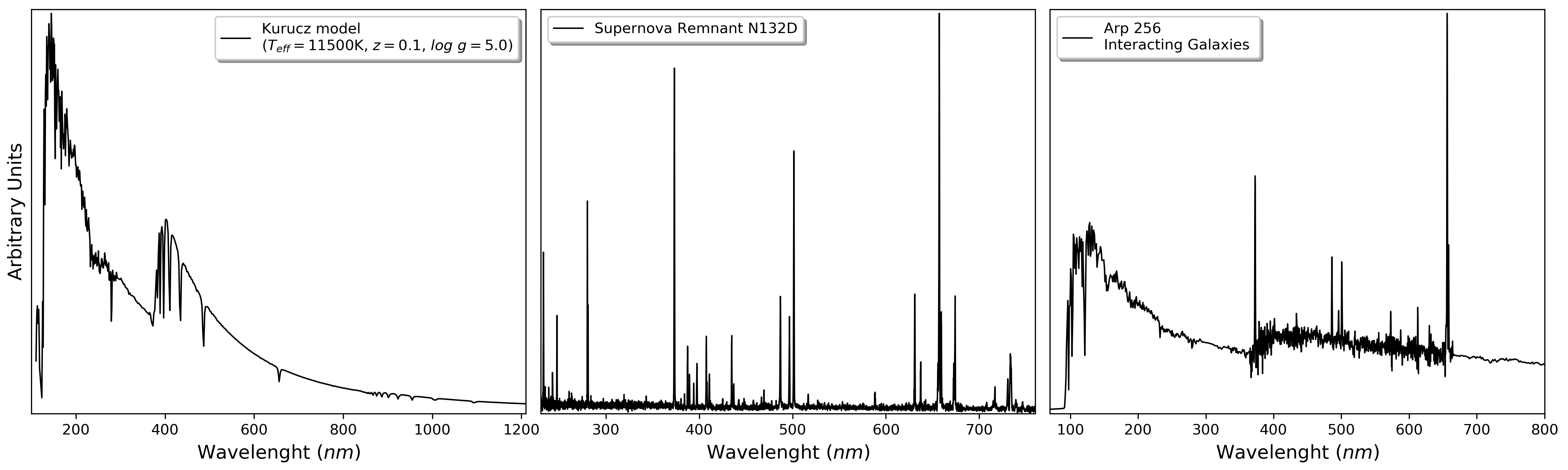}
	\caption{Example spectra used to compare the performance of the different algorithms: a Kurucz stellar atmosphere model (left), supernova remnant N132D in the Large Magellanic Cloud (center) and interacting galaxy Arp~256 (right). %The first and last spectra were generated with the ASTROLIB PYSYNPHOT \citep{Synphot} Python package and the middle one is a composition of five observations of the Faint Object Spectrograph (FOS) instrument of the Hubble Space Telescope (HST).
	}
	\label{fig:SpectraSample}
\end{figure*}
%----------------------------------------------------------------------------------
\subsubsection{LOWESS}
%----------------------------------------------------------------------------------

A popular variant of the Savitzky-Golay method is the locally weighted scatterplot smoothing (LOWESS) \citep{LOWESS}, where the regions to be fitted are not evenly spaced and the least squares procedure takes into account weighted values of the data, according to their distance from the point to be evaluated.
This scheme involves computing the coefficients of the fitted polynomial each time, producing a less efficient algorithm.
To compare with \fabada we use the implementation explained in \citep{LOWESS}, which uses a linear fit and can only be used for one dimensional data.
This implementation has only one parameter, which is the fraction $f$ of data points used to accomplish the linear regression at each point. For the optimisation we begin with the smallest fraction possible ($f=2/N_X$) and then it is incremented it by a factor of $1.5$ until the best recovery is found.

%----------------------------------------------------------------------------------
\subsubsection{Gaussian Filter (GF)}
%----------------------------------------------------------------------------------

Another classical technique of noise reduction consists in filtering the high frequency components of the data using a Gaussian filter.
The fast Fourier transform (FFT) is the most computationally efficient way to convert the data $z(x)$ to the frequency domain $\tilde z(w)$.
Once we have the spectrum of the image, defined as the amplitude of the FFT of the data, we can apply a low-pass Gaussian filter to discard the highest modes:
\begin{equation}
GF(\tilde z(w),W_0)=e^{-|\tilde z(w)|^{2} / 2 W_{0}^{2}}
\end{equation}
where $|\tilde z(w)|$ is the distance from the center (zero frequency component) in Fourier space, and $W_0$ is the radius of the filter, equivalent to $W_0 = 2\pi /R_0$ in configuration space.
% We can construct the smoothed data $\hat{y}_{GF}(x)$ by computing the inverse fast Fourier Transform of the filtered spectrum.
This is again a one-parameter method, in which we select the radius $R_0$ of the low-pass Gaussian filter by evaluating in twenty logarithmic steps between $R_0 = 1$ and $R_0 = 630$.

%----------------------------------------------------------------------------------
\subsubsection{Wiener Filter}
%----------------------------------------------------------------------------------

The Wiener filter minimizes the mean square error taking into account that the measurements are a random process where the statistical properties (in particular, the spectrum) of the noise are known.
In this work we have used the implementation in the \textit{Scipy Signal Tools} library \citep{SciPy,Wiener}, where the output of the filter given the signal $z(x)$ is given by
\begin{equation}
\hat{y}_W(x)=\left\{\begin{array}{cl}
\frac{\sigma^{2}}{\sigma_{z}^{2}} m_{z}+\left(1-\frac{\sigma^{2}}{\sigma_{z}^{2}}\right) z & \sigma_{z}^{2} \geq \sigma^{2} \\
m_{z} & \sigma_{z}^{2}<\sigma^{2}
\end{array}\right.    
\end{equation}
where $m_{z}$ and $\sigma_{z}^{2}$ are local estimates of the mean and variance within a window of size $w$, and $\sigma^2 = \langle\sigma_z^2(x)\rangle$ is the the average variance across the data.
We increment $w$ starting from 3 until the best recovery is found.

%----------------------------------------------------------------------------------
\new{\subsubsection{Total Variation filter (TV)}
The total variation filter is a well known algorithm, first described by \citet{TV}, that aims to minimise the image's total variation (TV) norm, defined as the square sum of the gradients of each pixel, in both directions.
In this work, we use the implementation of \citet{TV_Chambolle} in the \textit{scikit-image} library of Python that works either in one or two~dimensions.
This implementation has one parameter, the denoised~weight~$d_w$, and we increase it from $10^{-2}$ to $10^{3}$ in logarithmically spaced steps until the maximum PSNR is found. 
}

%----------------------------------------------------------------------------------
\new{\subsubsection{Wavelet denoising filter}
The Wavelet denoising filter is an adaptive approach to wavelet soft thresholding where a unique threshold is estimated for each wavelet subband. In this work, we used the implementation in the scikit-image library \citep{scikit-image}, while the method is described in \cite{Wavelet}. This algorithm is non-parametric, and no optimisation is necessary.
}

\begin{figure*}
	\includegraphics[width=\linewidth]{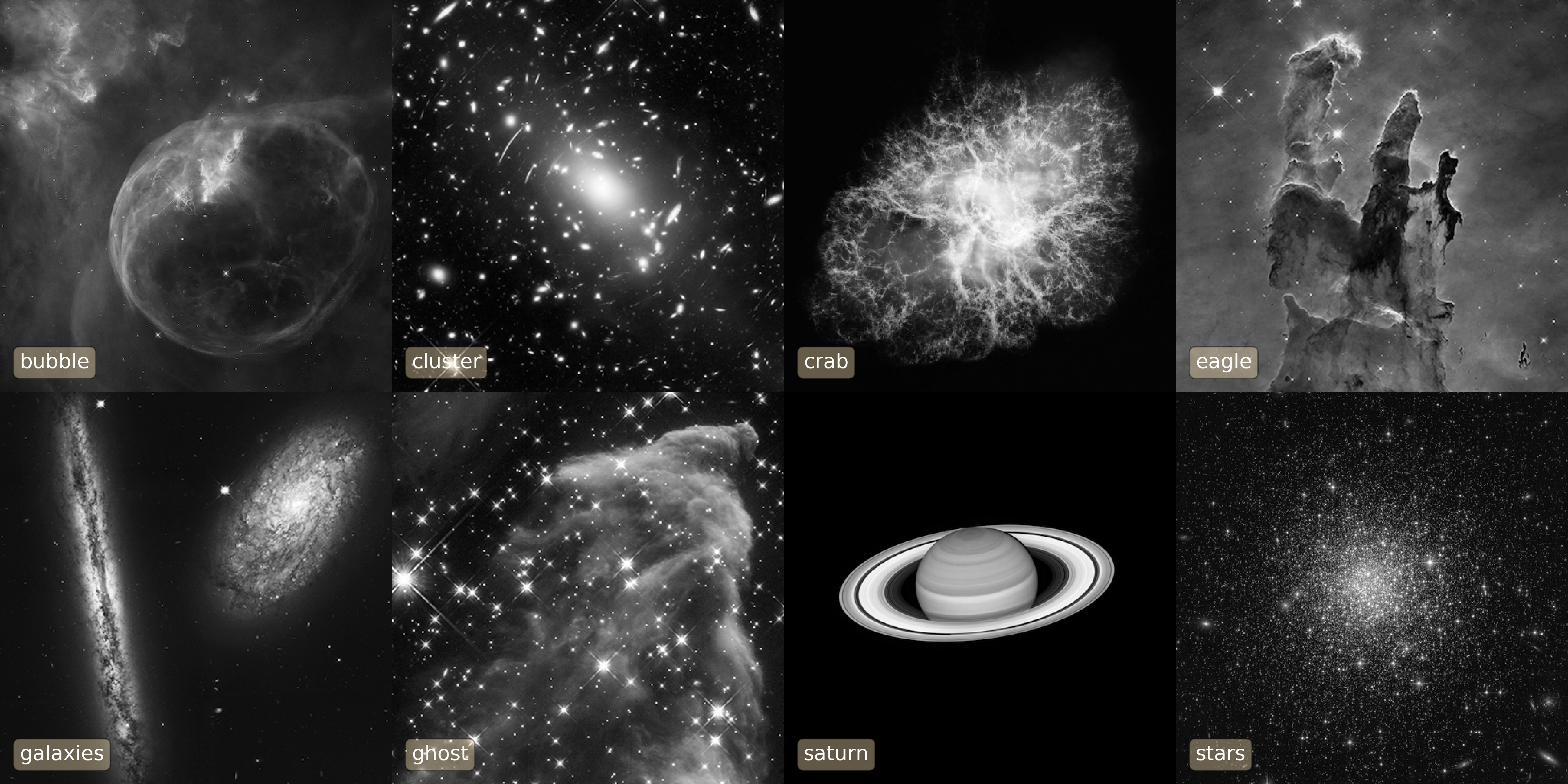}
	\caption{Battery of images used for the comparison procedure. From left to right, top to bottom, the objects shown in this figure are the Bubble nebula (NGC~7635), a galaxy cluster (Abell~S1063), the Crab nebula (M~1), the Eagle nebula (M~16), a spiral galaxy pair (NGC~4302 \&~4298), the Ghost nebula (IC~63), Saturn and a globular star cluster (NGC~1466). Labels on the bottom left corner are used to identify each object throughout this work.}
	\label{fig:ImagesSample}
\end{figure*}

%----------------------------------------------------------------------------------
\subsubsection{BM3D}
%----------------------------------------------------------------------------------

We also include the Block-Matching and 3D filtering \citep{BM3D} algorithm, which arguably represents the state of the art in the research field of image analysis \citep{CNN}.
A detailed account of this method, where image denoising is implemented as two-step process, can be found in \citep{BM3D}.
For the first step, the noisy image is divided into equal-size square blocks.
For each block, a 3D group is formed with similar regions (block-matching), and noise is attenuated by hard-thresholding the coefficients of a 3D transform.
The filtered image is used to estimate the energy spectrum of the signal, and the process can be repeated a second time using a Wiener filtering instead of hard thresholding.
The final smoothed result of the image is generated as a weighted average of the denoised blocks in their original positions.
This is a non-parametric method, and no optimisation is necessary.

%This approach yields a high-performance method for noise reduction, and converging in what is said, by some authors, the best analytical solution and little improvement in denoising performance were achieved until the advent of deep learning denoisers.

%----------------------------------------------------------------------------------
\new{\subsubsection{PySAP}
Finally, the  Python Sparse Data Analysis Package \citep[PySAP;][]{PySAP} is an open-source image processing software package developed
for the COmpressed Sensing for Magnetic resonance Imaging and Cosmology (COSMIC) project. This package provides
a set of flexible tools that can be applied to a variety of compressed sensing and image reconstruction problems. In particular, PySAP offers a denoising non-parametric automatic algorithm that is used in this work.
The denoising algorithm uses the isotropic undecimated wavelet transform from the C++ package, Sparse2D, to decompose the noisy image, and a soft threshold is then applied with weights learned from the noisy image itself. 
This is a non-parametric method, and no optimisation is necessary.
}

%----------------------------------------------------------------------------------
\subsection{Data Sample}
\label{sec:TestSample}

All the methods explained in the previous section are applied to a set of test data in one and two dimensions (astrophysical spectra and monochromatic images, respectively) with different levels of Gaussian random noise.

An important aspect in the recovery of spectra is the conservation of their features, such as the Balmer break or emission and absorption lines, after noise reduction.
For this purpose, we consider three different spectra (represented in Figure~\ref{fig:SpectraSample}) that show these characteristics in different degrees.
The first spectrum (left) is a Kurucz stellar atmosphere model \citep{Kurucz} with an effective temperature $T_{eff} = 11500 K$, metal abundance $\log Z=0.1$ and surface gravity $\log g=5.0$, typical of an O/B type star, with a prominent Balmer break at $\sim 400$~nm and several strong absorption lines.
The spectrum of a supernova remnant, plotted on the middle panel, is a composite of 5 different observations \citep{SN132D_Blair} from the Faint Object Spectrograph (FOS) instrument of the Hubble Space Telescope (HST).
This high-resolution spectrum ($0.9$ \AA/pixel) is characterized by very prominent emission lines, useful for inferring different physical properties of these objects.
The last spectrum (right) is taken from the TRDS Brown Atlas \citep{Arp256_Brown} which consist on a  pair of interacting galaxies, Arp~256, in the constellation of Cetus, and it contains a combination of emission and absorption lines with a stellar continuum. The Kurucz and Arp~256 spectra can be found in the Synthetic Photometry SYNPHOT \citep{Synphot} Python package that simulates photometric data and spectra, observed or otherwise. 
The aim of using these different spectra is to obtain a good representation of the possible features that can appear in one-dimensional astrophysical data and see how the different algorithms perform in digging up spectral features out of a noisy signal.

For astronomical images, we consider eight different targets, displayed in Figure~\ref{fig:ImagesSample}, that are intended to sample the wide range of features that may be encountered in the field, including planets, stars, diffuse nebulae, and galaxies, either alone or in potentially blended groups.
Saturn is arguably the target that is most similar to the ordinary test images (e.g natural landscapes, human subjects) that are often used in the context of digital image processing.
In addition, our sample includes two examples of nebulae (Crab and Bubble) dominated by the gaseous component, two with a more significant contribution of the stellar population (Eagle and Ghost nebulae), and a globular cluster full of stars with different brightness.
There is also an image with a galaxy pair (NGC~4302 \& NGC~4298) in which we can see two different orientations of the galaxies, as well as a galaxy cluster with a wide variety of morphologies and apparent sizes.
All of these images have been taken from the Hubble Space Telescope gallery produced by NASA and the Space Telescope Science Institute (STScI). All of them have been compressed to 8-bit values, with a maximal dynamical range of $0-255$ counts and $512\times 512$ pixels size to lighten up the computational load.
For simplicity, we have also normalized the astronomical spectra to $255$ in order to have the same dynamical range and represent the noise in terms of this value for both dimensions.

%----------------------------------------------------------------------------------
\subsection{Test Statistics}
\label{sec:TestStatistics}
%----------------------------------------------------------------------------------

We apply different levels of Gaussian random noise $\eta(x)$ with constant variance $\sigma^2$ to the real signal $y(x)$:
\begin{equation}
    z(x) = y(x) + \eta(x)
\end{equation}
where $\eta(x) = \mathcal{N}(0, \sigma)$, the element $x \in X$ denotes independent measurements (spectrum wavelengths or image pixels), and we assume that statistical errors are correctly characterised in the input data.
Once $z(x)$ is computed, a softened estimation $\hat{y}(x)$ of the real signal $y(x)$ is carried out using the different algorithms explained above.

In one dimension, noise levels vary from $\sigma = 5$ counts to $\sigma = 95$ counts, out of the $255$ maximal value that sets the dynamical range of our data (i.e. of the order of $\approx 2-40 \%$ relative errors).
We extend the values of $\sigma$ to even higher values in two-dimensional data, specifically to $1024$ counts ($400\%$) to illustrate the challenging situation (not so seldom encountered in astronomy) that the signal is actually fainter than the background noise.

We evaluate the quality of the reconstruction in terms of the Peak~Signal~to~Noise~Ratio~(PSNR) and the Structure Similarity Index Measure (SSIM) of the estimators $\hat{y}(x)$, following common practice in the signal processing literature.

By definition, the PSNR (usually expressed in decibels, dB) is related to the Mean Square Error (MSE)
\begin{equation}
\mathrm{MSE}(\hat{y})=\frac{1}{N_X} \sum_{x \in X}\left(\hat{y}(x)-y(x)\right)^{2}
\label{eq:MSE}
\end{equation}
as
\begin{equation}
\mathrm{PSNR}(\hat{y}) = 10 \cdot \log _{10}\left(\frac{255^{2}}{\mathrm{MSE}}\right),
\label{eq:PSNR}
\end{equation}
where  $255$ is the dynamical range in our data.
In principle, a more faithful recovery of the underlying signal should yield smaller values of the MSE and higher values of the PSNR.

\begin{figure*}
    \includegraphics[width=\linewidth]{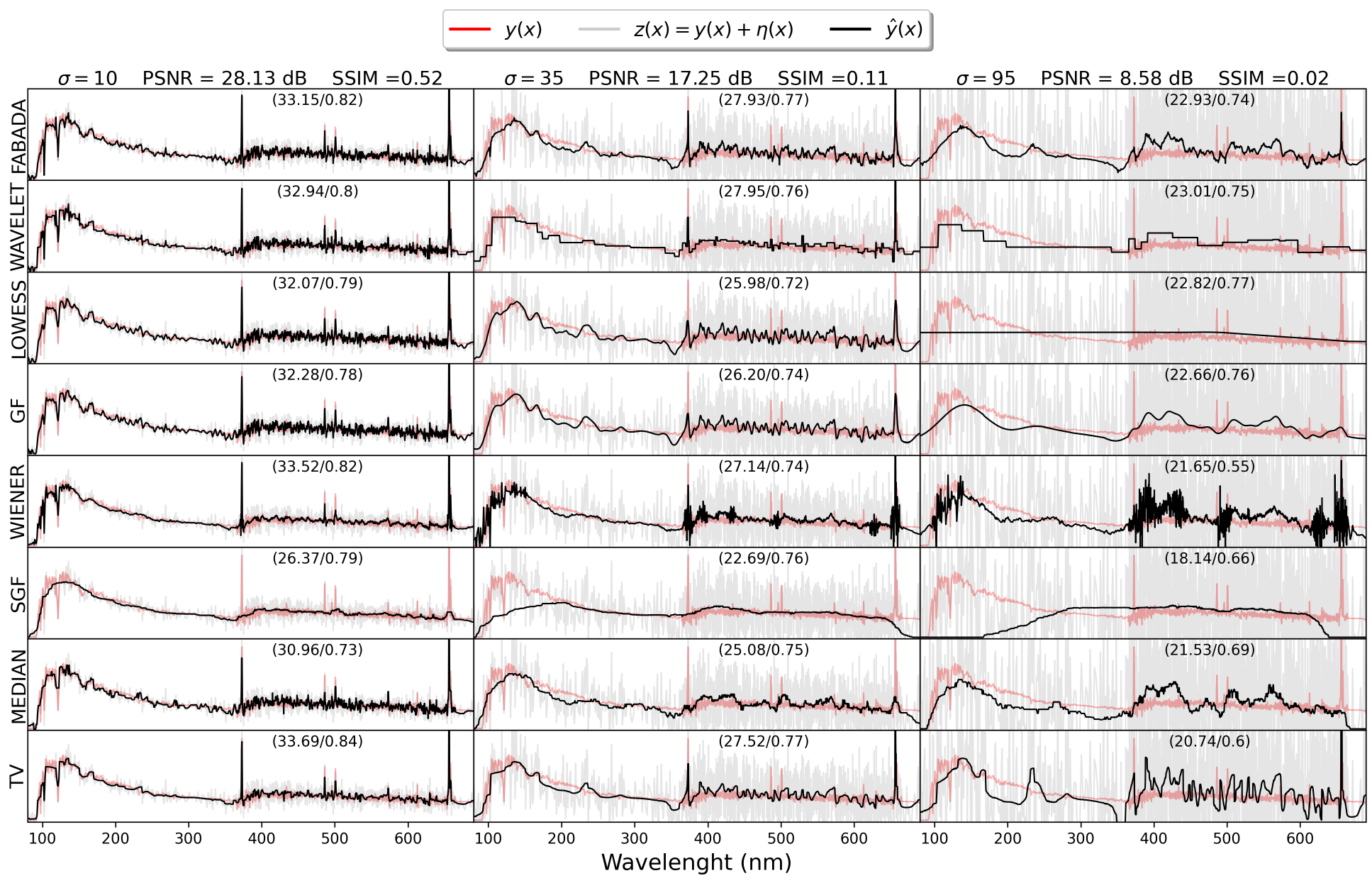}
\caption{\new{Results obtained for the Arp~256 spectrum (see Figure~\ref{fig:SpectraSample}) by all the models explained in Section~\ref{sec:OtherAlgoritms} (rows) for three different noise levels (from left to right columns).
The real signal $y(x)$, the noisy input data $z(x)$, and the estimation $\hat{y}(x)$ are represented as red, grey and black lines, respectively.
Numbers on each panel quote the PSNR and SSIM obtained by each method, to be compared with the noisy data (column headers).}}
\label{fig:RecoverSample_1D}
\end{figure*}

The Structural Similarity Index Measure (SSIM) is another typical metric used in image restoration that takes into account properties such as luminance, contrast and structure.
It is defined by the expression \citep{SSIM}
\begin{equation}
\operatorname{SSIM}(\mathbf{x}, \mathbf{y})=\frac{\left(2 \mu_{x} \mu_{y}+C_{1}\right)\left(2 \sigma_{x y}+C_{2}\right)}{\left(\mu_{x}^{2}+\mu_{y}^{2}+C_{1}\right)\left(\sigma_{x}^{2}+\sigma_{y}^{2}+C_{2}\right)}
\label{eq:SSIM}
\end{equation}
where $x$ and $y$ denote the two images being compared, $\mu$ and $\sigma^2$ are their mean and variance, and $\sigma_{xy}$ their covariance.
The constants $C_1$ and $C_2$ are two variables to stabilize the division when the denominator approaches zero, and they are usually set to $C_1 = (0.01\, L)^2$ and $C_2 = (0.03\, L)^2$, where $L$ is the dynamic range of the images. The SSIM metric can adopt values from 0 (absolute lack of correlation) to 1 (high structural similarity).

%It is very important to note here that, as can be readily seen in table \ref{tab:table2}, only BM3D and \fabada are completely parameter-free.
%In all other cases, the parameters of each algorithm have been optimised to to minimize the MSE (maximize the PSNR) of each particular realization of the input data.
%This implies that their results are the best possible and should be regarded as an upper limit to the performance of these algorithms, since this kind of optimisation is only possible when the correct solution is known.

Another metric that we consider is the CPU time used to generate the estimation of the real data on a 2.40 GHz Intel i9-9980HK CPU along with 16Gb DDR4 2400 MHz RAM memory.
Please note that this time corresponds to the final execution time for the given noise level in the Python implementation of the algorithms, once the optimal parameters have been found, but it does \emph{not} include the time invested in the optimisation, which is considerably larger.

\begin{figure*}
    \includegraphics[width=0.95\linewidth]{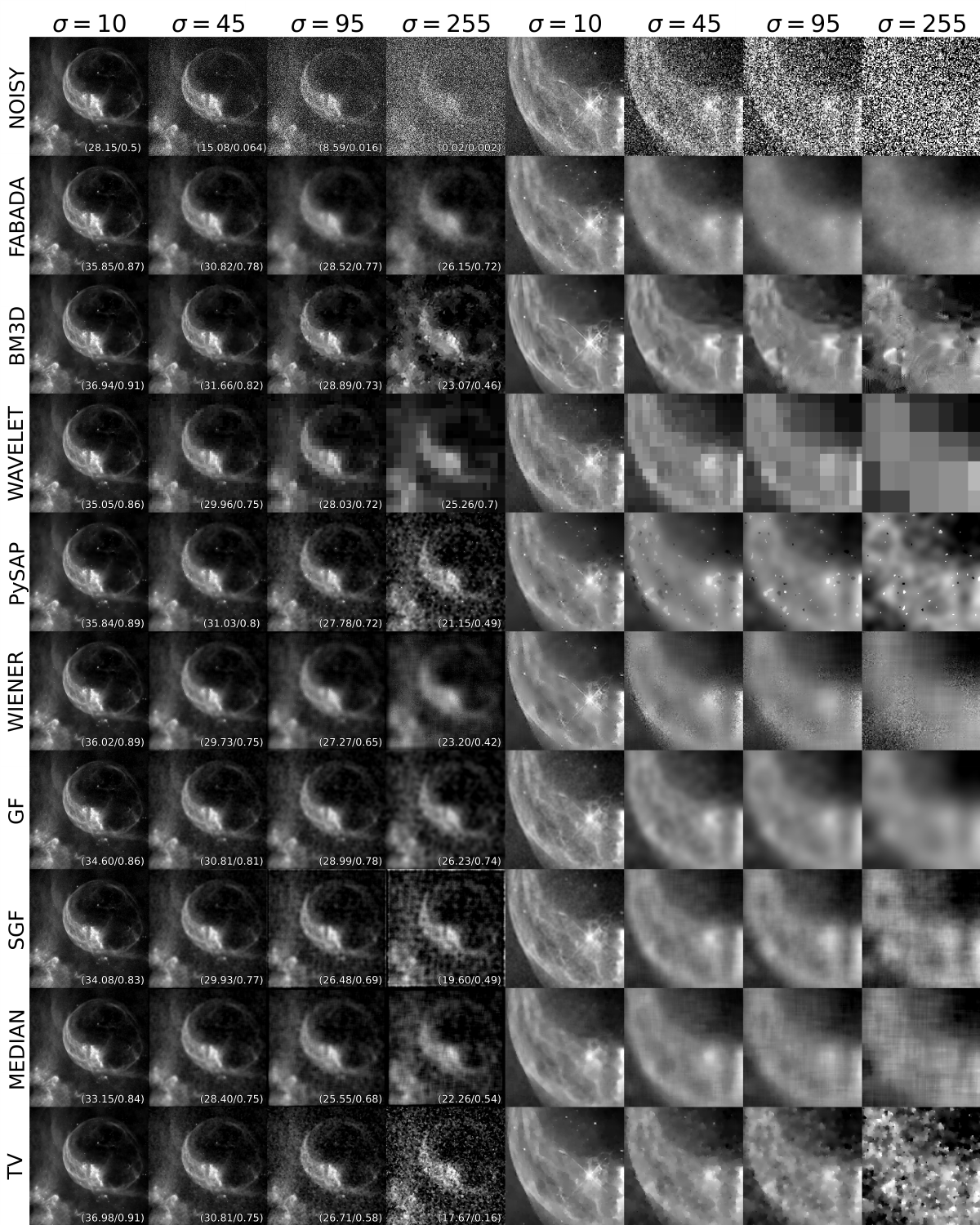}%height=0.95\textheight
\caption{\new{Example of the estimation results for the bubble image (see Figure~\ref{fig:ImagesSample}), showing the recoveries for three different values of noise ($\sigma=10,45,95,255$), i.e. values of PSNR of $28.15$, $15.08$, $8.59$, and $0.02$~dB and SSIM of $0.5$, $0.064$, $0.016$, $0.002$). Rows and columns represent different algorithms and noise levels, respectively, as indicated by the labels. The last three columns show a close-up of the bubble structure to better gauge the recovery of fine details.}} %\vspace{1cm}}
\label{fig:RecoverSample_2D}
\end{figure*}

%----------------------------------------------------------------------------------
\section{Results}
\label{sec:Results}
%---------------------------------------------------------------------
We now assess the ability of our algorithm to recover the underlying signal for the synthetic test cases described in Section~\ref{sec:TestSample}.
In order to facilitate the comparison with previous results reported in the literature, we use the Peak Signal-to-Noise Ratio (PSNR) defined in~\eqref{eq:PSNR}, which is just a measure of the Mean Squared Error (MSE), expressed in decibel (dB), as well as the Structural Similarity Index Measure (SSIM) defined in~\eqref{eq:SSIM}. All the results shown from the parametric methods are optimise using the PSNR metric (similar results are obtained if SSIM was used instead). 

\subsection{Recovery Examples}

Figures~\ref{fig:RecoverSample_1D} and~\ref{fig:RecoverSample_2D} show two examples of the results obtained by the different algorithms explained in Section~\ref{sec:OtherAlgoritms}: the median, Savitzky-Golay filter (SGF), Gaussian Filter (GF), Wiener filter, and locally weighted scatterplot smoothing (LOWESS) for one-dimensional spectra, and the median, SGF, GF, Wiener filter and block-matching (BM3D) filters for the images.
Each row of both figures provides the recovered estimations of the signal $\hat{y}(x)$ for different noise levels, represented in each column.
\subsubsection{Spectra Example}
We represent in figure~\ref{fig:RecoverSample_1D} the recoveries obtained for three random realizations with high, low, and extremely low signal-to-noise ratios (SNR) of the \textit{Arp~256} spectrum (see Figure~\ref{fig:SpectraSample}).
The PSNR achieved by each method is quoted within the corresponding panel along with the SSIM index.

At high SNR ($\sigma=10$, original PSNR$=28.13$~dB, SSIM$=0.52$), all algorithms display not only a similar performance, but actually converge to very similar solutions.
\new{The highest value of PSNR/SSIM is obtained by the optimised TV filter (33.69~dB / 0.84), a little better than the Wiener filter (33.52~dB / 0.82) and \fabada (33.15~dB / 0.82).}
The difference is lower than the difference between these three recoveries and any of the others.
The improvement with respect to the originally high-quality data is necessarily modest in all cases, of the order of $\approx 3$ dB, i.e. an increase of $\sim 50\%$ in MSE.
All algorithms are able to correctly trace the presence of the most prominent emission lines, as well as the strong Lyman-$\alpha$ absorption line near the peak at the left end of the spectrum.
Nevertheless, it is important to note that, while the \new{TV filter provides the best recovery in terms of overall noise reduction, \fabada tends to preserve the true intensity of these features slightly better than any of the other algorithms.}

This trend becomes more significant as the noise increases, and it is more difficult to discriminate significant spectral features from Gaussian random fluctuations.
In the middle panels, where $\sigma = 35$, all models are able to reproduce the overall shape of the continuum.
However, they fail to recover even the strongest absorption and emission lines, although hints of the brightest emission lines are still present in the \new{Wavelet}, LOWESS, Wiener,  GF and TV filters.
Only our prescription is able to provide a good description of these prominent features with this level of noise in the input data, albeit weaker absorption and emission lines are completely lost.
This reflects on the values of the metrics, where \fabada and the Wavelet filter obtain the highest values ($27.93$~dB/$0.77$) and ($27.95$~dB/$0.76$) respectively; a difference of $10.68$~dB, which is more than an order of magnitude of noise reduction with respect to the original measurements. However, the \new{Wavelet filter seems to have a decreased resolution since it merges similar regions in wider bins}. 

If we now turn to the recovery of the noisiest spectrum ($\sigma=95$, on the right column), \new{we see this behaviour of the wavelet filter more noticeable. Despite obtaining the highest values of the metrics, almost all the information enclosed in the spectrum is gone.} In comparison, \fabada obtains the second highest value of PSNR, while GF and LOWESS obtain higher values of SSIM and close values of PSNR. While the results of \fabada and GF are actually similar, all the spectral information in LOWESS is gone.
At these high levels of noise, the MSE and the SSIM metrics are rather inadequate to assess the quality of the reconstructed emission line spectra, because they incur in minimal penalty for failing to reproduce a handful of peaks that are barely statistically significant.
The criteria implemented in \fabada are more conservative, and a lot of random fluctuations are kept (hence the slightly lower SSIM) together with the most significant remains of the actual signal.
It is somewhat remarkable that \fabada manages to recover the brightest line even at this noise level, at variance with the other methods, while still obtaining high values of the noise reduction metrics.
We stress once again that this does not necessarily imply a failure of the methods, but of the MSE as a goodness-of-fit indicator.
On the other hand, it does highlight the robustness of \fabada in this respect, although the results reported here suggest that perhaps the MSE and SSIM is not the optimal metric to gauge the quality of the recovered solution or, more likely, that it should be complemented with another test statistic that quantifies information loss and/or gives more weight to informative features.

% ----------- 2D Example ----------------
\subsubsection{Image Example}
Similar trends are observed in the results obtained for the 2D data.
Figure~\ref{fig:RecoverSample_2D} shows the recovery of the \textit{Bubble} nebula image for the whole set of different algorithms, represented in each row, \new{with four different noise levels ($\sigma=10,\ 45,\ 95,\ 255$)} along the columns.
The PSNR and SSIM values obtained for each are also shown in each panel, and a close-up of some structures is provided in the last four columns to illustrate whether the smoothing methods can reproduce their shape and edges.
All models yield fairly similar reconstructions for the highest signal-to-noise case ($\sigma=10$).
The best reconstruction in terms of the MSE and SSIM is provided by the \new{TV method, which improves the PSNR from $28.15$~dB to $36.98$~dB, more than one order of magnitude of noise reduction in terms of the MSE, and from $0.5$ to $0.91$ for the SSIM index. The state-of-the-art  BM3D obtains similar results with $36.94$~dB and SSIM of $0.91$.}
This is around $15 \%$ more than the recovery with \fabada ($35.85$~dB) in terms of PSNR and $10\%$ ($0.87$) in terms of SSIM. Similar results are obtained with the Wiener filter ($36.02$~dB/$0.89$) and \new{with PySAP ($35.84$~dB/$0.89$)}.
The other classical filters (median, SGF, GF) obtain comparable results when their parameters are tuned to minimise the MSE, about $\approx 1$~dB ($\sim 25\%$) below \fabada's solution.
Regardless of the MSE and SSIM statistics, the Wiener filter and \new{PySAP virtually miss some of the filaments in the shape of the zoomed structure,} while the other classical filters recover some of the shapes, although in a blurrier way than \fabada, TV and BM3D.
Visually, the \new{TV filter} and BM3D reproduce the shape of the zoomed structure better, both in terms of sharpness and smoothness.
In particular, several edges in \fabada seem to be a little bit more blurry compared to BM3D, together with a clearly visible \textit{salt and pepper} noise component.

\new{Similar behaviour is found when we increase the noise to $\sigma = 45$. \fabada yields results ($30.82$~dB/$0.78$) comparable to the best solutions obtained by BM3D ($31.66$~dB/$0.82$) and PySAP ($31.03$~dB/$0.8$). Despite its lower PSNR ($30.81$~dB), the GF obtains a higher value of SSIM ($0.82$) than \fabada. If we inspect the zoomed region, we see that PySAP is able to effectively filter the high-frequency noise with the cost of not recovering well the gradients and the structure. In contrast, BM3D is able to recover fairly well the gradients of the image, while \fabada and GF recover a blurrier image.}

\begin{table*}
%\resizebox{\textwidth}{!}{%
\begin{tabular}{|l|c|c|c|c|l|c|c|c|c|c|c|c|c|}
\hline
\multicolumn{5}{|l|}{\hspace{2.9cm}1D} & \multicolumn{9}{l|}{\hspace{4.9cm}2D} \\ \hline
\multirow{3}{*}{Method} & \multicolumn{2}{c|}{PSNR (dB)} & \multicolumn{2}{c|}{SSIM} & \multirow{3}{*}{Method} & \multicolumn{4}{c|}{PSNR (dB)} & \multicolumn{4}{c|}{SSIM} \\ \cline{2-5} \cline{7-14} 
 & \multicolumn{2}{c|}{all $\sigma$} & \multicolumn{2}{c|}{all $\sigma$} &  & \multicolumn{2}{c|}{all $\sigma$} & \multicolumn{2}{c|}{$\sigma > 95 $} & \multicolumn{2}{c|}{all $\sigma$} & \multicolumn{2}{c|}{$\sigma > 95 $} \\ \cline{2-5} \cline{7-14} 
 & $\#_{max}$ & $\delta$ & $\#_{max}$ & $\delta$ &  & $\#_{max}$ & $\delta$ & $\#_{max}$ & $\delta$ & $\#_{max}$ & $\delta$ & $\#_{max}$ & $\delta$ \\ \hline
\fabada & 10 & \textbf{0.575} & 5 & 0.118 & \fabada & 6 & \textbf{1.430} & 6 & \textbf{0.576} & 4 & 0.116 & 1 & 0.113 \\ \hline 
Wavelet & 15 & 0.598 & 4 & 0.065 & Wavelet & 17 & 1.776 & 17 & 0.653 & 15 & \textbf{0.102} & 15 & \textbf{0.07} \\ \hline 
LOWESS & 0 & 1.205 & 2 & 0.071 & BM3D & \textbf{60} & 1.436 & 10 & 2.858 & \textbf{54} & 0.112 & 8 & 0.216 \\ \hline 
Savitzky–Golay & 4 & 1.377 & 9 & 0.083 & Savitzky–Golay & 0 & 4.439 & 0 & 5.361 & 3 & 0.19 & 3 & 0.219 \\ \hline 
Gaussian & 9 & 0.885 & \textbf{23} & \textbf{0.033} & Gaussian & 23 & 1.54 & \textbf{22} & 0.598 & 25 & 0.117 & \textbf{22} & 0.081 \\ \hline 
Median & 0 & 1.866 & 0 & 0.112 & Median & 1 & 3.17 & 1 & 2.337 & 4 & 0.191 & 4 & 0.199 \\ \hline 
Wiener & \textbf{18} & 0.615 & 13 & 0.129 & Wiener & 0 & 5.421 & 0 & 7.864 & 2 & 0.192 & 2 & 0.257 \\ \hline 
TV & 1 & 2.613 & 1 & 0.243 & TV & 5 & 4.247 & 0 & 7.168 & 3 & 0.204 & 0 & 0.322\\ \hline 
- & - & - & - & - & PySAP & 0 & 3.897 & 0 & 4.848 & 2 & 0.162 & 1 & 0.22\\ \hline 

\end{tabular}
%}
\caption{Table showing the number of tries that a specific algorithm has recovered the best solution ($\#_{max}$) for each metric (PSNR, SSIM). Along these values, it is shown the average euclidean distance to the maximum value, which we defined as \new{$\delta = <PSNR_{max} (\sigma) - PSNR_i(\sigma)>$  for each algorithm $i$}.}
\label{tab:metric_dist}
\end{table*}

Once again, the advantages of our algorithm become more evident as the noise increases.
In the middle range, with a noise level of $\sigma=95$, one may see that \fabada's difference in MSE with respect to BM3D decrease to $0.37$~dB, \new{and it becomes better than PySAP by $0.7$~dB, achieving higher values of SSIM in both cases}.
On the other hand, the GF reaches the highest values of MSE and SSIM, whereas the Wiener, \new{TV}, SGF and median methods display increasingly lower values than the other methods.
Most of them are able to bring up large-scale gradients and structures, but they fail to recover the smallest filaments (as shown by the zoomed panels) due to the high level of noise.
The most significant problem of \fabada is still the salt and pepper noise.
\new{The PySAP denoising filter, in turn, returns a composition of smoothed and high-contrast regions that hardly reproduces the true underlying gradients in the signal.
The Wavelet filter again seems to merge regions with similar intensity, yielding a lower effective resolution in the final recovery.
Finally},
%ABAJO: This again gives us a hint of the MSE and SSIM metrics may not be the optimal way to quantify the recovery of this type of image.
%On the other hand,
it is remarkable how the state-of-the-art BM3D method starts to add some artificial edges to the recovered image.

If we push the noise even further, as is often the case in practical astrophysical applications, the signal itself is comparable to or even lower than the statistical uncertainties.
In our last test, the noise level is $\sigma = 255$, equal to the dynamic range of the original data.
The GF obtains again the highest values of MSE ($26.23$~dB, $0.08$~dB higher than \fabada's solution), almost an order of magnitude of noise reduction, and an SSIM value of 0.74, 0.02 point higher than \fabada.
\new{Despite the artificial lower resolution, the Wavelet filter is able to recover high values of the metrics (25.26~dB/ 0.7), 
which again gives us a hint that the MSE and SSIM metrics may not be the optimal way to quantify the recovery of this type of image}.
It is noteworthy to see that these three solutions are far above the rest, by more than $\approx 3$~dB in PSNR and $\approx 0.3$ in SSIM.

Upon visual inspection of the different reconstructions, including the zoomed region, one can see that almost every algorithm, except \fabada and GF, produces some artefacts in the images either at large or small scales.
The GF, \new{however}, is very efficient at filtering high-frequency noise, but, by construction, highlights graininess at the characteristic scale of the filter.
A similar effect is \new{more evident in the PySAP, TV, SGF, and median filters}.
\new{Although the wavelet filter is able to recover the large-scale structure of the image, the decrease of resolution is even more noticeable, and spurious futures associated with the wavelet basis appear in the reconstruction}.
The Wiener filter still mixes regions with very different levels of smoothing, while BM3D is extremely successful in eliminating the high-frequency 'grain', albeit the smooth areas of the real image are transformed into more staggered gradients.
This feature might be inherent to the block-matching algorithm, whose aim is to classify similar square sections of the image into groups, thus resulting in patches with similar gradients and/or edges.
The fully automatic method developed in this work, \fabada, seems to offer a reasonable compromise solution.
At high levels of noise, it is able to reduce the MSE and increase the SSIM at the cost of significantly blurring the image, but the introduction of spurious patterns, apart from salt and pepper noise, is not as severe as in the other methods.
% This discussion becomes even more relevant in the context of astrophysical data where the consequences of adding these artificial features may alter the physical conclusions. 

\begin{figure*}
   \includegraphics[width=1\linewidth]{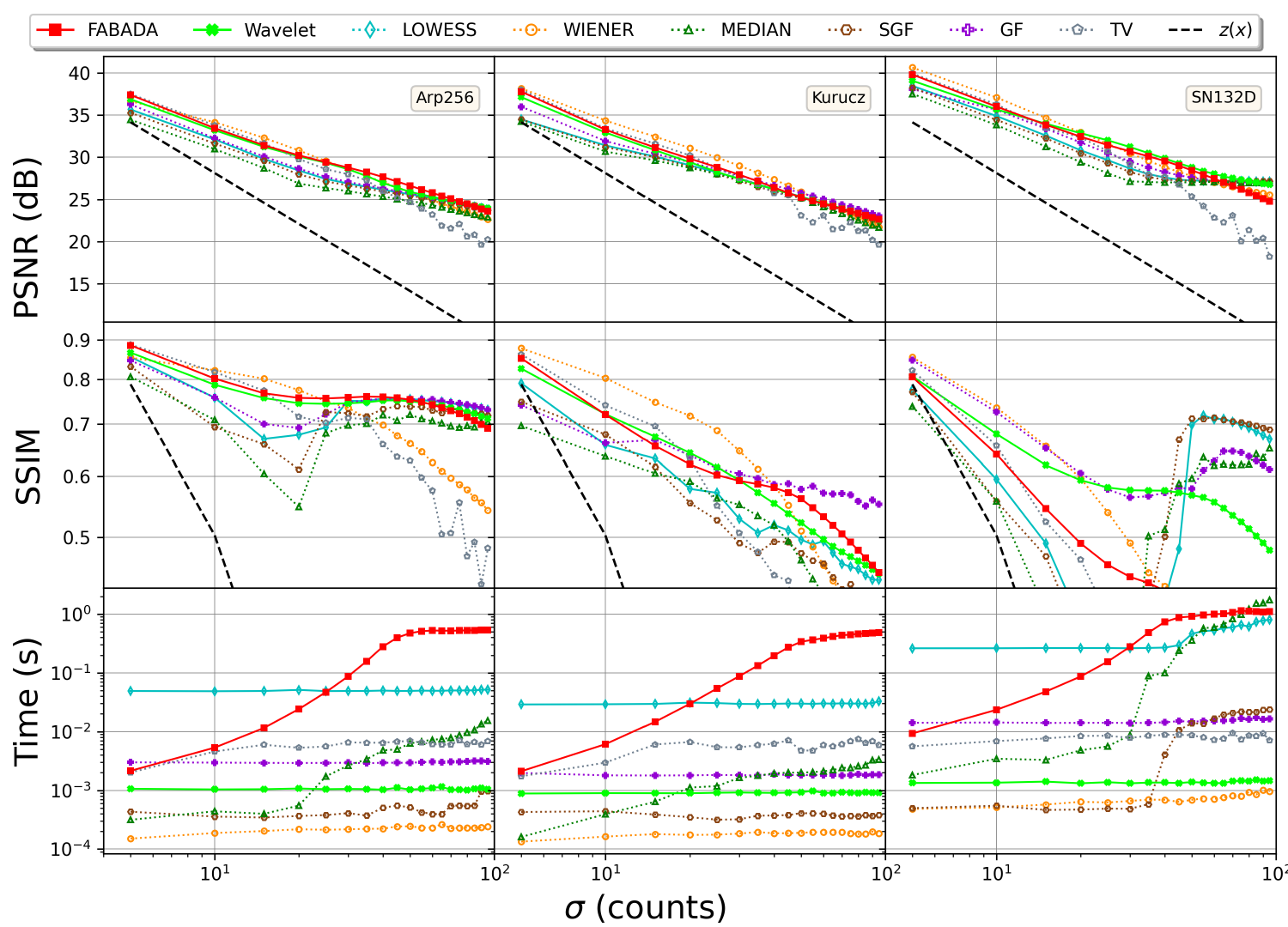}
   \caption{Peak Signal to Noise Ratios (PSNR), Similarity Structural Indexes (SSIM), and CPU times obtained for all the one-dimensional spectra samples and noise ranges considered in the comparison procedure. In the top set of figures is shown the PSNR in decibels~(dB), in the middle is represented the SSIM, and in the bottom one the CPU time in seconds~(s). Each figure of both groups is labelled with the reference name given at the top of the column. The dashed yellow line represents the PSNR and SSIM of the noisy data. Solid lines with filled symbols refer to the non-parametric methods. In 1D data, \new{we used two automatic methods}, the one presented in this work, \fabada, which is represented with a red solid square, \new{and the Wavelet filter, with light green x}. Dotted lines with unfilled symbols refer to the optimised methods. The LOWESS algorithm is represented with the blue diamond, the Wiener filter with the orange circle, the median filter with the green triangle, SGF with the brown hexagon, the  GF with the purple cross, and \new{the TV filter with the grey pentagon}.}
\label{fig:Results_1D}
\end{figure*}

\begin{figure*}
   \includegraphics[width=0.9\linewidth]{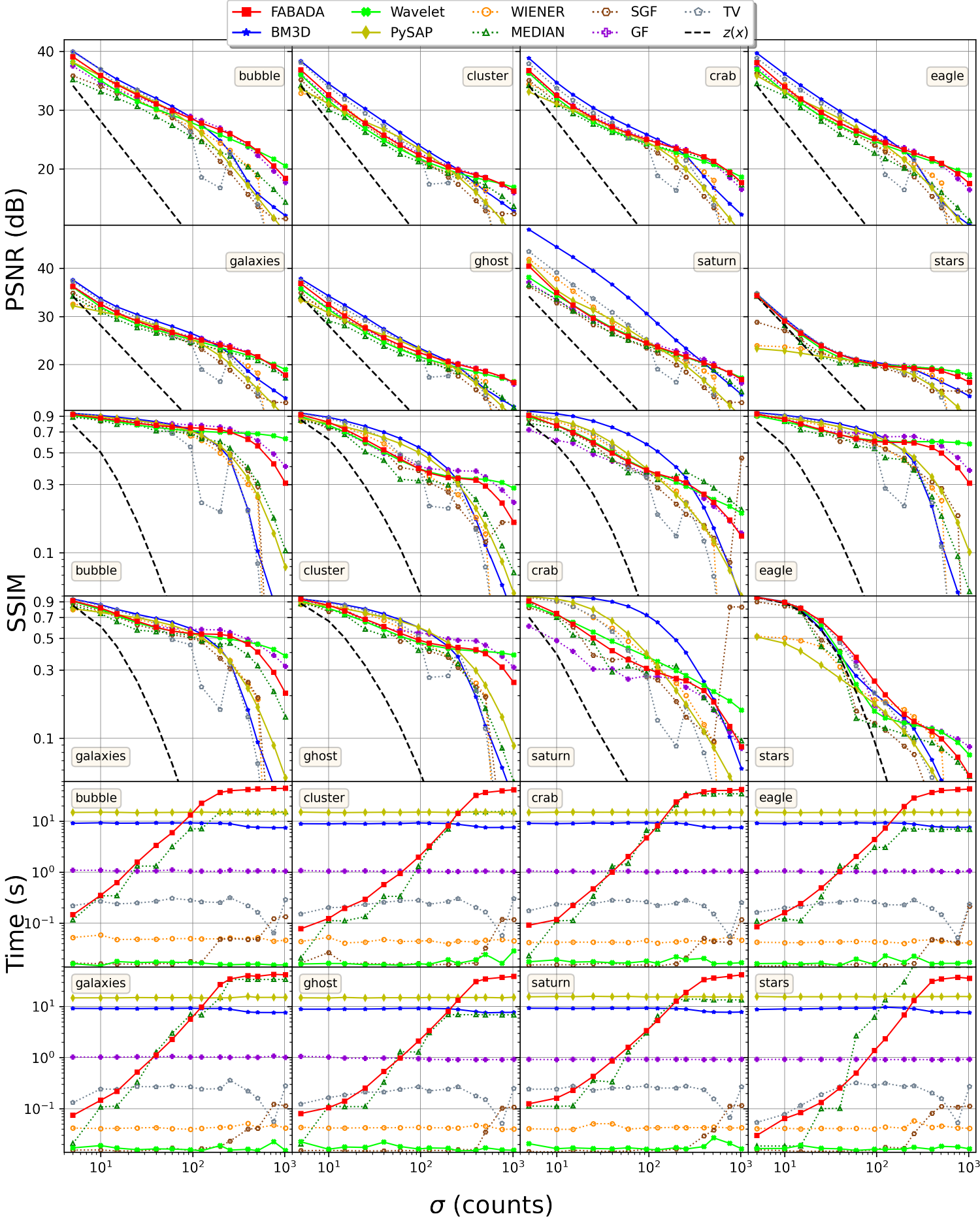}
   \caption{Peak Signal to Noise Ratios (PSNR), Similarity Structural Indexes (SSIM), and CPU times obtained for all the two-dimensional image samples and noise ranges considered in the comparison procedure. In the top set of figures is shown the PSNR in decibels~(dB), in the middle is represented the SSIM, and in the bottom one is the CPU time in seconds~(s). Each panel is labelled with the reference image name given. The dashed yellow line represents the PSNR and SSIM of the noisy data. Solid lines with filled symbols refer to the non-parametric methods. In 2D data, there are four automatic methods, \fabada, which is represented with a red solid square, the block matching 3D filtering BM3D, with blue stars, \new{the Wavelet filter with light green exes, and the PySAP with dark yellow diamonds}. Dotted lines with unfilled symbols refer to the optimised methods. The Wiener filter is represented with the orange circle, the median filter with the green triangle, SGF with the brown hexagon, GF with the purple cross and \new{the TV filter with the grey pentagon}.}
\label{fig:Results_2D}
\end{figure*}

%%------------------------------------
%%------ALL RESULTS DISCUSSION---------
%%------------------------------------

\subsection{Entire database results}

We have computed the PSNR, SSIM, and CPU time for all data samples, noise levels, and methods as a function of $\sigma$, averaged over ten different random realisations.
The number of cases that a specific algorithm has recovered the best solution, as well as the average difference with respect to the optimal choice, are quoted in Table~\ref{tab:metric_dist}.

\subsubsection{One dimension - Spectra}
In one dimension (Figure~\ref{fig:Results_1D}), \fabada behaves very similarly to the optimised Wiener and the Wavelet filters, particularly in terms of PSNR (top panel).
In general, the best results regarding this metric are achieved with the latter methods; 18 out of the 57 ($32\%$) test cases by the Wiener filter, followed by the \new{Wavelet filter with 15 ($26\%$)} and \fabada, who obtained the best PSNR for 10 ($18\%$) estimations, being especially successful in the low signal-to-noise regime.
Then, the optimised GF obtained the best reconstruction in 9 ($16\%$) cases, and the remaining 5 ($10\%$) correspond to the optimised SGF (4) and the \new{TV filter (1)}.
Neither LOWESS nor the median filter obtained in any case the best values for PSNR.
Just taking into account these results, one can see how \fabada performs as well as the best possible solutions of the standard methods typically used in astronomy.
The average difference with respect to the highest PSNR achieved by any algorithm is only $0.575$~dB, a little closer than the \new{Wavelet method $0.598$~dB}.
This supports the idea that \fabada automatically achieves the limit of the standard methods when their parameters are tuned to the (unknown) optimal values. 

In terms of the SSIM metric (middle panels), our algorithm obtains slightly worse results, being the best option only in $5$ ($9\%$) test cases, with an average distance with respect to the highest SSIM of $0.118$.
The optimised GF filter obtained $23$ ($40\%$) of the highest values and an average difference of $0.034$.
This result is due in part to the $SN132D$ spectra, where most of the information is concentrated in the emission lines.
At high values of the noise level, almost all standard methods converge towards a horizontal line, where all the information is lost (see e.g. LOWESS in Figure~\ref{fig:RecoverSample_1D}).
However, the SSIM (and, to some extent, PSNR) do not strongly penalise the misfitting of the emission lines when their statistical significance becomes low, which hints the limitations of these metrics to quantify the recovery of relevant information.
Discarding the SN132D spectra, \fabada is seldom the optimal choice, but the average distance decreases to $0.048$ in terms of SSIM, compared to $0.022$ for the GF and $0.056-0.085$ for the other algorithms.

As regards to the CPU time, it is easy to see that \fabada has a strong dependence on the level of noise, at variance with most classical methods, due to the increasing number of iterations required to fulfil our $\chi^2$ stopping condition.
This behaviour is also seen in the median, due to a similar increase in the optimal window size, whether other methods are less sensitive to the noise level.
\fabada is, in general, significantly slower than the other algorithms (except LOWESS, at high signal-to-noise ratios), although it must be borne in mind that we have not taken into account the time consumed by the optimisation process of the standard algorithms, only the execution time once the optimal parameters have been found.
%The fastest methods are the Wiener Filter and SGF, with an average time of $0.0004$ and $0.003$ seconds respectively. This is due to its simple implementation where they only have to carry out some light calculations. The GF and the LOWESS, with average times of $0.08$ and $0.22$ seconds, respectively, have bigger variance between the spectra data sample due to the difference in the length of the data. While \textit{SN132D} had around $8000$ data points, \textit{Arp 256} had $2300$ data points, and the \textit{Kurucz} model only had around $1000$. Beside the strong dependence on \fabada, we can still compute the average time which results on $0.467$ seconds, three orders of magnitude slower than the fastest algorithm, the Wiener filter. If we divide the sample in two difference regimes, high and low SNR, $\sigma<45$ and $\sigma \ge 45$, we found that \fabada obtains averages times of $0.123$ and $0.752$ seconds, respectively. Despite of \fabada being  the slowest algorithm for almost both regimes (LOWESS is slower at high SNR) in practice, we would have to spend time to select the parameter for the rest of algorithm, and we might not even choose the best ones. This advantage causes that the overall time need to use the standards algorithms slower than \fabada, in practice.  
\subsubsection{Two dimensions - Images}
For the two-dimensional images (Figure~\ref{fig:Results_2D}) BM3D reaches the highest PSNR for $60$ out of our $112$ ($54\%$) test cases with different target and noise levels, followed by the optimised GF with $23$ ($21\%$), the \new{Wavelet filter $17$ ($15\%$)}, \fabada with $6$ ($5\%$),  \new{TV filter with $5$ ($4\%$)} and $1$ ($1\%$) that correspond to the optimised median filter.
Neither SGF, \new{PySAP}, nor the Wiener filter ever obtained the highest values.
Nevertheless, BM3D is on average $1.44$~dB below the optimal choice, comparable to the $1.43$~dB of \fabada, $1.50$~dB of GF, and \new{$1.77$~dB of the Wavelet filter} due to the significantly different trends observed at different noise levels.

In general, BM3D stands over the other methods, including \fabada, at high SNR ($\sigma \lesssim 95$~dB), in particular for the \textit{Saturn} image.
Its collaborative filter is particularly well suited for periodic data, or images with repetitive patterns, which are virtually absent in other test cases.
The \textit{stars} image would be a paradigmatic example, and the difference in this test is insignificant.

On the other hand, \fabada, \new{the Wavelet filter}, and GF predominate in the low-SNR regime ($\sigma \gtrsim 95$~dB), where BM3D only achieves the best reconstruction in $10$ out of the $56$ ($18\%$) test cases.
The remaining $46$ are achieved by GF with $22$ ($39\%$), \new{the Wavelet filter with $17$ ($30\%$)}, \fabada with $6$ ($12\%$), and the median filter with $1$ ($1\%$).
Furthermore, BM3D is on average $2.86$~dB below the highest value, while \fabada,  GF, and \new{the Wavelet filter are only $0.57$, $0.59$, and $0.65$~dB lower, respectively.}
Essentially, \fabada, \new{the Wavelet filter}, and the optimised GF achieve remarkably similar results, and they perform better than BM3D at low signal-to-noise ratios. \new{Despite the high values obtained by the Wavelet filter, we see in our examples that small-scale details in the image are lost, and they become replaced by artificial structures arising from the wavelet basis. This is not seen either in \fabada or BM3D.}

This behaviour is also seen in the middle panels, where the SSIM is represented.
From the $54$ ($48\%$) of the highest values achieved by BM3D, only $8$ ($19\%$) are above $\sigma > 95$ counts, and the difference with respect to the highest value of SSIM increases with noise level from $0.1$ to $0.2$.
Although in this metric \fabada only achieves the best recovery in $4$ ($5\%$) cases, $1$ ($7\%$) at high noise levels, it is only $0.1$ apart, on average, from the best solution.
The optimised GF obtains the highest solution for $23$ ($15\%$) cases and $25$ ($13\%$) in the low SNR regime.
\new{Additionally, the Wavelet filter produces better estimations than \fabada in $17$ ($15\%$) cases, and $15$ ($13\%$) in the low SNR regime.}
Despite this difference, \new{GF and the Wavelet filter are still $0.11$ and $0.12$ below the best solution ($0.07$ and $0.081$ at low SNR) respectively, again very close to the results obtained by our algorithm.}

This change in the trend of the best solution is at the cost of CPU time.
While \fabada seems to converge to its solution in an efficient way in the high SNR regime (consistently below the second), for high noise levels (low SNR) the time rises up to $30$~s, as shown in the bottom panels.
Once again, for the optimised models we only take into account the execution time with the fine-tuned parameters already known.
The results are similar to the one-dimensional case, but the time scale increases with the problem's dimensionality.
\new{The Wavelet filter is the fastest algorithm, obtaining an average time of $0.02$ seconds, while the PySAP denoising filter seems to be the slowest, with $18.24$ seconds.}
\fabada also features a high execution time ($8.08$ seconds on average), whereas BM3D, GF, \new{TV,} the Wiener filter, and SGF yield $8.7$, $0.87$, \new{$0.05$} and $0.04$ seconds, respectively.
However, the performance of \fabada and the median filter vary significantly as a function of SNR.
Combining these results with the above metrics for the quality of the reconstruction, we argue that our method provides a competitive alternative both at high SNR, where it achieves slightly less reliable results than the state-of-the-art algorithm BM3D at a fraction of the computational cost, as well as in the low SNR regime, where it provides a more faithful reconstruction after a significantly larger execution time.

\section{Conclusions}
\label{sec:Conclusions}

In this work, we present the theory and implementation of a novel automatic algorithm for noise reduction: the Fully Adaptive Bayesian Algorithm for Data Analysis (\fabada).
Our method iteratively evaluates progressively smoother models of the underlying signal and then combines them according to their Bayesian evidence and $\chi^2$ statistic.
The source code is publicly available at \url{https://github.com/PabloMSanAla/fabada}.

We compare \fabada with other methods that are representative of the current state of the art in image analysis and digital signal processing.
For this comparison, we used the most typical metrics, the Peak-Signal-to-Noise-Ratio (PSNR), which is a measure of the Mean Square Error (MSE), the Structural Similarity Index (SSIM), and the CPU time.
One important advantage of our method, shared by BM3D, PySAP, and the Wavelet filter, over classical algorithms is the absence of free parameters to be tuned by the user.
Our results suggest that \fabada, the Wavelet filter and BM3D achieve values of PSNR or SSIM comparable to or better than the best possible solution attainable by the classical methods.
%\new{Discuss well about wavelet filter that decreases considerably the spatial resolution introducing artificial pixels, although obtains really high values of the metric}
Beyond the precise values of the global quantitative metrics, both \fabada and BM3D are quite successful in adapting to the structures present in the input data.
Perhaps the most significant difference between them is that \fabada's priors assume that the signal is smooth, whereas BM3D uses block-matching to look for repetitive patterns.
This might be relevant when one must recover the height and shape of the spectral features in 1D or the gradients and boundaries in 2D.
We argue that \fabada appears to offer a trade-off between noise reduction, increasing the metric values significantly, in a way that is statistically compatible with the data, keeping significant features without introducing considerable artifacts.

Regarding execution time, non-parametric methods are more expensive than the classical alternatives \emph{once the optimal values of their parameters are known}, something that is of course impossible in practice.
\fabada is faster than BM3D at high SNR, although it usually yields a poorer reconstruction, and the trend reverses for low SNR.

\section{Data availability}
The Python implementation of the method developed in this work is publicly available at \url{https://github.com/PabloMSanAla/fabada}. The images used in this work were taken from the Hubble Space Telescope gallery produced by NASA and the Space Telescope Science Institute (STScI). The spectra used in this work were taken from \cite{Kurucz,SN132D_Blair,Arp256_Brown}. 

\section*{Acknowledgements}
We acknowledge financial support from the State Research Agency (AEI-MCINN) of the Spanish Ministry of Science and Innovation under the grant "The structure and evolution of galaxies and their central regions" with reference PID2019-105602GB-I00/10.13039/501100011033, from the ACIISI, Consejería de Economía, Conocimiento y Empleo del Gobierno de Canarias and the European Regional Development Fund (ERDF) under grant with reference PROID2021010044, and from IAC project P/300724, financed by the Ministry of Science and Innovation, through the State Budget and by the Canary Islands Department of Economy, Knowledge and Employment, through the Regional Budget of the Autonomous Community.
Yago Ascasibar acknowledges financial support from grant "Starbursts throughout the evolution of the Universe" (PID2019-107408GB-C42/AEI/10.13039/501100011033) from the AEI-MICINN, Spain.

\bibliographystyle{rasti}
\bibliography{Bibliography}

\bsp	% typesetting comment
\label{lastpage}
\end{document}